\documentclass[a4paper,12pt]{article}
\usepackage{amssymb,amsfonts,amsmath,mathtext,amsthm,cite,enumerate,float, authblk}

\usepackage{color}      

\title{Towards a tomographic representation of quantum mechanics on the plane}

\author[1, 2]{G.G. Amosov\thanks {gramos@mi.ras.ru}}

\author[2]{A.I. Dnestryan\thanks {dnestor@inbox.ru}}
 
\affil[1]{Steklov Mathematical Institute of RAS}
\affil[2]{Moscow Institute of Physics and Technology}

\begin{document}

\maketitle

\begin{abstract}
On the base of symplectic quantum tomogram we define a probability distribution on the plane $\mathbb {R}^2$.
The dual map transfers all observables which are polynomials of the position and momentum operators
to the set of polynomials of two variables. In this representation the average values of
observables can be calculated by means of integration over all the plane.
\end{abstract}

\section{Introduction}
In the twentieth century there were several attempts to describe quantum mechanics not with help of operators, but by the language of functions, as in classical mechanics. For this it is necessary to assign a certain (maybe quasi) probability distribution function to the quantum state, and another function defined on the same space to the operator of observable. In other words, let $\hat \rho$ and $\hat a$ be a density operator (a positive Hermitian operator with unit trace) and an observable, respectvely. Then, these operators should be associated with the function $\rho(X)$ and $a(X)$,
\begin{equation}\label{1}\hat \rho \leftrightarrow\rho(X), \;\;\;\;\; \hat a\leftrightarrow a(X), \;\;\;\; X\in(\mathcal X, \mu),\end{equation}
on the measurable space $(\mathcal X,\mu)$,
wherein the average value of $\hat a$ in the state $\hat \rho$ can be calculated by the formula
\begin{equation}\label{srednee}
<\hat a>_{\hat \rho} = \int_{\chi} \rho(X)a(X)d\mu(X).
\end{equation} 
The realisation of this program resulted in the requirement of taking the appropriate set of test functions $\{\rho (X)\}$ for
which $\{a(X)\}$ are distributions (generalised functions).

In 1932, Wigner \cite{Wigner} introduced the function, later named in his honor
$$W(q,p)=\frac{1}{2\pi}\int_{\mathbb{R}}\rho\Big(q+\frac{u}{2},q-\frac{u}{2}\Big)e^{-ipu}du$$
where $\rho(\cdot,\cdot)$ is the density matrix of $\hat \rho $ in the coordinate representation. Wigner noticed that for functions either only of the position operator $\hat q$ or of the momentum operator $\hat p$ alternatively, the average values can be calculated according to the formulas
$$<F(\hat q)>_{\hat \rho}=\int_{\mathbb{R}^2}F(q)W(q,p)dqdp$$
$$<G(\hat p)>_{\hat \rho}=\int_{\mathbb{R}^2}G(p)W(q,p)dqdp$$
\par
In 1949 Moyal \cite{Moyal} generalized this result to the case of arbitrary polynomials of  $\hat q$ and $\hat p$.
He showed that the mathematical expectation of the symmetrized polynomial of the form
\begin{equation}\label{sym}
\{\hat q^m \hat p^n\}_s = \sum_{s=0} ^nC_n^s \hat p^s \hat q^m \hat p^{n-s}
\end{equation}
can be calculated by the formula
$$<\{\hat q^m \hat p^n\}_s>_{\hat \rho}=\int_{\mathbb{R}^2}F_{mn}(q,p)W(q,p)dqdp$$
at that $F_{mn}(q,p)=q^mp^n$. Thereby Moyal proposed method of calculating average values for a broad class of observables through the Wigner function along the lines of classical mechanics. However, it should be noted that the Wigner function $W(q,p)$ is not positive in general.  So it can not be regarded as a classical probability distribution function. 
\par
In \cite{Bertrand, Vogel} the optical quantum tomogram ${\bf w}(X, \varphi)$ was introduced, defined as the Radon transform of the Wigner function
\begin{equation}\label{opt}
{\bf w}(X,\varphi)=\int_{\mathbb{R}^2}W(q,p)\delta(X-q\cos\varphi -p\sin\varphi)dqdp.
\end{equation}
Function ${\bf w}(X,\varphi)$ is non-negative everywhere and a one-parameter set $$\{{\bf w}(X,\varphi), \ \varphi\in[0,2\pi)\}$$ consists of a probability distributions on the real axis.

In \cite {MMT} it was noticed that the optical tomogram ${\bf w}_{\hat \rho}(X,\varphi)$ of a quantum state $\rho$ can be calculated directly from the density operator
$${\bf w}_{\hat \rho}(X,\varphi)=\text{Tr}\big(\hat\rho\delta(X\hat I-\hat q\cos\varphi -\hat p\sin\varphi)\big).$$
Similarly there was introduced a more general notion of symplectic quantum tomogram
\begin{equation}\label{sympl}
\omega _{\hat \rho}(X,\mu, \nu)=\text{Tr}\big(\hat\rho\delta(X\hat I-\mu\hat q -\nu\hat p)\big).
\end{equation}
Symplectic tomogram has the property of homogeneity
\begin{equation}\label{hom} 
\omega (\lambda X,\lambda\mu, \lambda\nu)=\frac{1}{|\lambda|}\omega (X,\mu, \nu)
\end{equation}
and it contains the same information as a density operator. The tomographic representation of quantum mechanics was developed in many branches
during elapsed time \cite {Ibort, Marg}. Such new approaches as $C^*$-algebras formalism appeared to be connected with this topic \cite {Ibort2}.

In \cite{Amo1, Amo2} it was studied the following question. Suppose that one has taken for the correspondence
$\hat \rho \to \rho (X)$ in (\ref{1}) the map from the set of states with sufficiently smooth density matrices $\rho (\cdot ,\cdot )$
in the coordinate representation to the set of optical tomograms $w(X,\varphi)$ (in the case, variable $X\equiv (X,\varphi )$). 
Then, the dual map $\hat a\to a(X)$ in (\ref {1}) is given by the extension of
\begin{equation}\label{dual}
a(X,\phi )=-2\pi \lim \limits _{\varepsilon\to 0}Tr(\hat a(X+i\varepsilon -\cos (\phi )\hat q-\sin (\phi )\hat p)^{-2})
\end{equation}
from the set of density operators $\hat a$.
It was shown that for the symmetrized product of canonical quantum observables $\{\hat q^m\hat p^n\}_s$ defined in
(\ref {sym}) the dual map results in
\begin{equation}\label{symb}
a_{mn}(X,\varphi)=X^{m+n}H_{m,n}(\varphi)
\end{equation}
where functions $H_{m,n}(\varphi)$ are biorthogonal to the system $\{\cos^{n+m-k}\varphi \sin^k \varphi\}$, i. e.
\begin{equation}\int\limits_0^{2\pi} \cos^{n+m-k}(\varphi )\sin^k (\varphi)\; H_{n+m-s,s}(\varphi)d\varphi =\delta_{ks}.
\end{equation}
\par
In the present paper we pass from the polar coordinates $(X,\varphi)$ to the Cartesian coordinates $(x,y)$. Our aim is to introduce a probability distribution $\Omega (x,y)$ 
on the plane  $\mathbb {R}^2$ such
that the average values can be calculated by the formula
\begin{equation}\label{main}
<\{\hat q^m \hat p^n\}_s>_{\hat \rho}=\int\limits_{-\infty}^{+\infty}\int\limits_{-\infty}^{+\infty}f_{\hat a}(x,y)\Omega(x,y)dxdy
\end{equation}
where function $f_{\hat a}(x,y)$ corresponds to $a_{mn}(X,\varphi)$ in the Cartesian coordinates.

\section{The comparison with the traditional approach}

To calculate the average value of the observable $\hat a$ in the state $\hat \rho $ in the traditional representation of quantum mechanics one should
take a trace \cite{Neu}
\begin{equation}\label{sred}
<\hat a>_{\hat \rho}=Tr(\hat \rho \hat a).
\end{equation}
In this section we discuss the advantages of applying (\ref {srednee}) instead of the traditional approach (\ref {sred}). 

First and foremost, the optical tomogram (\ref {opt}) is a measurement of the homodyne quadrature \cite {Smit, Sch}. After one obtains it, the density
operator $\hat \rho $ should be reconstructed from the experimental data. If we consider the tomogram as the main unit describing the quantum system,
instead of the density operator, then this reconstruction is not needed. It allows one to take maximum advantage of experimental
data. Such an approach results in a number of significant tasks concerning the situation in which one should reconstruct a tomogram in the case of
incomplete information \cite{1, 2, 3, 4}.  

On the other hand, the tomographical technique can be successfully used for comparing classical and quantum systems in the sense of \cite {CQ}.
Indeed,  the role of a density operator in the classical statistical mechanics is played by a probability distribution on the phase plane. Therefore, to compare
classical and quantum cases probability distributions are preferable to density operators.
In this framework, one can discriminate what kind of probability distributions on the plane refer either to the classical or the quantum case. It is worth mentioning that 
using the symplectic tomogram allows one to take a limit for the centre-of-mass tomogram \cite {centre} under the condition that a number of particles tends to
infinity \cite {limit}. Surely these are but several examples of the possibilities of using tomographic picture.

\section{Tomography on the plane}

In \cite {Amo1, Amo2} the optical tomogram ${\bf w}(X,\phi )$ (\ref {opt}) plays a role of a probability distribution on the set ${\mathbb R}\times [0,2\pi ]$.
Indeed, 
\begin{equation}\label{dobava2}
{\bf w}(X,\varphi )\ge 0,\ \frac {1}{2\pi }\int \limits _0^{2\pi }\int \limits _{-\infty }^{+\infty }{\bf w}(X,\varphi )dXd\phi =1.
\end{equation}
In turn, the information containing in the set $\{{\bf w}(X,\varphi )\}$ is redundant due to that
\begin{equation}\label{dobava}
{\bf w}(-X,\varphi)={\bf w}(X,\varphi +\pi)
\end{equation}
(it immediately follows from (\ref {opt})). It results in the knowledge of $\{{\bf w}(X,\phi ),\ X\ge 0,\ \phi \in [0,2\pi]\}$ allowing reconstruction of 
the entire tomogram.
The pair $(X,\phi )\in {\mathbb R}_+\times [0,2\pi ]$ can be considered as the polar coordinates on the plane ${\mathbb R}^2$.
Thus, it looks natural to change the polar coordinate to the Cartesian coordinate. This should give a transformation of 
${\bf w}(X,\phi )$ to a probability distribution $\Omega (x,y)$ on the plane ${\mathbb R}^2$.
Let us define a function $\Omega(x,y)$ in terms of the symplectic tomogram (\ref {sympl}) by the formula (here we used the homogeneity (\ref {hom}))
\begin{equation}\label{plane}\Omega(x,y) = \frac{1}{\sqrt{x^2+y^2}}\omega \bigg(\sqrt{x^2+y^2}, \frac{x}{\sqrt{x^2+y^2}}, \frac{y}{\sqrt{x^2+y^2}}\bigg) \end{equation}
$$
= \omega (x^2+y^2,x,y).
$$
It follows then that
\begin{equation}\label{dobava3}
\Omega (x,y)\ge 0,\ \frac {1}{\pi}\int \limits _{{\mathbb R}^2}\Omega (x,y)dxdy=1.
\end{equation}
Moreover, let us assume that $a(X,\phi)$ is the symbol of observable $\hat a$ in the optical tomography representation (\ref {dual}). 
Let us presume that
$$
f_{\hat a}(X\cos (\phi),X\sin(\phi ))=a(X,\phi ).
$$

\underline{Assertion 1}:
$$
<\hat a>_{\hat \rho}=\int \limits _{\mathbb {R}^2}f_{\hat a}(x,y)\Omega (x,y)dxdy.
$$

\underline{Proof}. 
Consider the right-hand side of (\ref {main}) and pass to the polar coordinates
$$\int\limits_{-\infty}^{+\infty}\int\limits_{-\infty}^{+\infty}f_{\hat a}(x,y)\Omega(x,y)dxdy=$$
$$=\int\limits_0^{2\pi}\int\limits_{0}^{+\infty}f_{\hat a}(r\cos\varphi,r\sin\varphi)\Omega(r\cos\varphi,r\sin\varphi)\frac{\partial(x,y)}{\partial(r,\varphi)}drd\varphi=$$
$$
=
\int\limits_0^{2\pi}\int\limits_{0}^{+\infty}f_{\hat a}(r\cos\varphi,r\sin\varphi)\frac{1}{r}\omega (r, \cos\varphi, \sin\varphi)rdrd\varphi=
$$
$$
\int\limits_0^{2\pi}\int\limits_{0}^{+\infty}f_{\hat a}(r\cos\varphi,r\sin\varphi){\bf w}(r,\varphi)drd\varphi.
$$
As required.

Put
\begin{equation}\label{s}
f_{mn}(X\cos\varphi,X\sin\varphi)=a_{mn}(X,\varphi),
\end{equation}
where $a_{mn}$ is defined in (\ref {symb}). The function $f_{mn}(x,y)$ determined by (\ref {s}) is a symbol for 
the observable $\hat a=\{\hat q^m\hat p^n\}_s$.

Let us consider the Hilbert space consisting of all polynomials of $N$th degree on the plane $\mathbb {R}^2$ equipped with the scalar product
\begin{equation}\label{scalar}
(f,g)_N=\frac {2}{N!}\int \limits _{{\mathbb R}^2}\exp \left (-x^2-y^2\right )f(x,y)\overline g(x,y)dxdy.
\end{equation}

\underline{Assertion 2}: The functions $(f_{mn}(x,y))_{m+n=N}$ form the biorthogonal system to the functions $(x^{k}y^{N-k})_{k=0}^N$
with respect to the scalar product (\ref {scalar}).

\underline{Proof}. 

Passing to the polar coordinate we obtain that $f(X,\Phi)=X^NF(\phi)$ and  $g(X,\Phi)=X^NG(\phi)$ are
biorthogonal iff $F(\phi )$ and $G(\phi)$ are bioorthogonal on the unit circle. 

In the Appendix we find the functions $f_{mn}$ for $m+n=2$ and $m+n=3$. Thus,
$$
f_{20}(x,y)=\frac {1}{2\pi }(3x^2-y^2),\ f_{11}(x,y)=\frac {4}{\pi }xy,\ f_{02}(x,y)=\frac {1}{2\pi }(3y^2-x^2),
$$
\begin{equation}\label{funkcii}
f_{30}(x,y)=\frac {2}{\pi}(x^3-xy^2), f_{21}(x,y)=\frac {2}{\pi}(5x^2y-y^3), 
\end{equation}
$$
f_{12}(x,y)=\frac {2}{\pi}(5xy^2-x^3),\ f_{03}(x,y)=\frac {2}{\pi}(y^3-x^2y).
$$

\section{Some simple examples}

Let us calculate $\Omega (x,y)$ determined by (\ref {plane}) for coherent states. The symplectic quantum tomogram of
a coherent state $|\alpha >$ with the wave function
$$
<X|\alpha >=\frac {1}{\pi ^{1/4}}\exp\left (-\frac {X^2}{2}+\sqrt 2\alpha X-\frac {\alpha ^2}{2}-\frac {|\alpha |^2}{2}\right )
$$
is known to be equal to (see, f.e., \cite {coherent})
$$
\omega _{\alpha}(X,\mu ,\nu)=\frac {1}{\sqrt {\pi (\mu ^2+\nu ^2)}}\exp\left (-\frac {(X-\sqrt 2Re(\alpha )\mu-\sqrt 2Im(\alpha )\nu)^2}{\mu ^2+\nu ^2}\right ).
$$
Applying (\ref {plane}) we obtain
$$
\Omega _{\alpha }(x,y)=\frac {1}{\sqrt {\pi (x^2+y^2)}}\exp\left (-\frac {(x^2+y^2-\sqrt 2Re(\alpha )x-\sqrt 2Im(\alpha )y)^2}{x^2+y^2}\right ).
$$
In particular, for a vacuum state $|0>$ it implies
$$
\Omega _{0}(x,y)=\frac {1}{\sqrt {\pi (x^2+y^2)}}\exp(-x^2-y^2).
$$
The number of particles operator $\hat N=\frac {\hat q^2+\hat p^2-1}{2}$ has the symbol
$$
f_{\hat N}(x,y)=\frac {1}{2}(f_{20}(x,y)+f_{0,2}(x,y)-1)=\frac {1}{2\pi }(x^2+y^2-1)
$$
due to (\ref {funkcii}). It immediately follows that
$$
<\hat N>_0=\int \limits _{\mathbb {R}^2}\Omega _0(x,y)f_{\hat N}(x,y)dxdy=0.
$$
Analogously, for the excited states $|n>$ of harmonic oscillator, with the wave functions 
$$
<X|n>=\frac {1}{\pi ^{1/4}}\frac {1}{\sqrt {2^nn!}}H_n(X)\exp \left (-\frac {X^2}{2}\right ),
$$
where $H_n(X)$ is the Hermitian polynomial of nth degree,
the symplectic tomogram equals
$$
\omega _n(x,\mu ,\nu)=\frac {1}{2^nn!}\frac {1}{\sqrt {\pi (\mu ^2+\nu ^2)}}H_n^2\left (\frac {X}{\sqrt {\pi (\mu ^2+\nu ^2)}}\right )\exp \left (-\frac {X^2}{\mu ^2+\nu^2}\right ).
$$
Then, it follows from (\ref {plane}) that
$$
\Omega _n(x,y)=\frac {1}{2^nn!}\frac {1}{\sqrt {\pi (x^2+y ^2)}}H_n^2\left (\sqrt {\frac {x^2+y^2}{\pi}}\right )\exp (-x^2-y^2).
$$
The same reasoning gives us
$$
<\hat N>_n=\int \limits _{\mathbb {R}^2}\Omega _n(x,y)f_{\hat N}(x,y)dxdy=n.
$$

\section{Conclusion}

Based upon the symplectic quantum tomogram we introduced the probability distribution on the plane (\ref {plane}). Then, we have presented
the dual map allowing to calculate the average values of quantum observables in the quantum state determined by this distribution.
Finally, we have calculated the symbols of quantum observables which are polynomials of the position and momentum operators of
second and third degrees. Few simple examples of calculating the probability distribution on the plane and the symbols of observables are
given. We suppose that our approach can serve for solving the task of reconstructing a tomogram
from incomplete information about the system. We shall treat with this problem in the future.

\section*{Acknowledgments} The work is fulfilled under the support of Russian Scientific Foundation (Project N 14-21-00162).

\section*{Appendix}

\par
Here we find a biorthogonal system $f^{(n)}_k(x,y)$ to the auxiliary system\\ $g^{(n)}_s(x,y)=x^{n-s}y^s$ with respect to the scalar product
$$(f,g)=\frac{2}{n!}\int\int e^{-x^2-y^2}f^{(n)}_k(x,y)\overline g^{(n)}_s(x,y)dxdy$$
\\
\par
Obviously $f^{(1)}_1(x,y)=\frac {x}{ \pi },\; f^{(1)}_2(x,y)=\frac {y}{ \pi}$. Look for now $f^{(2)}_k(x,y)$.
We have:
$$g^{(2)}_1(x,y)=x^2, \;\;g^{(2)}_2(x,y)=xy, \;\;g^{(2)}_3(x,y)=y^2$$
Functions $f^{(2)}_k(x,y)$ are polynomials of the second degree relative $x$ and $y$, therefore they represent a linear combination of functions $g^{(2)}_s(x,y)$ that is
$$f^{(2)}_k(x,y)=a_kx^2+b_kxy+c_ky^2$$
While the basis is
 $g^{(2)}_s(x,y)$ we can write the column coordinate functions $f^{(2)}_k(x,y)$ as $\begin{pmatrix}a_k\\b_k\\c_k\end{pmatrix} $. To calculate the scalar products $(f,g)$
we write the Gram matrix in the basis of $g^{(2)}_s(x,y)$:
$$\Gamma = \begin{pmatrix} (g^{(2)}_1,g^{(2)}_1) &(g^{(2)}_1,g^{(2)}_2) &(g^{(2)}_1,g^{(2)}_3) \\
 (g^{(2)}_2,g^{(2)}_1) &(g^{(2)}_2,g^{(2)}_2) &(g^{(2)}_2,g^{(2)}_3) \\
 (g^{(2)}_3,g^{(2)}_1) &(g^{(2)}_3,g^{(2)}_2) &(g^{(2)}_3,g^{(2)}_3)
\end{pmatrix}$$
where 
$$(g^{(2)}_i,g^{(2)}_j)=\int\int e^{-x^2-y^2}g^{(2)}_i(x,y)g^{(2)}_j(x,y)dxdy=\int\limits^{+\infty}_{-\infty} x^{4-i-j}e^{-x^2}dx\int\limits^{+\infty}_{-\infty} y^{i+j}e^{-y^2}dy$$
The values of these integrals are known, so
$$\Gamma = \frac{\pi }{4}\begin{pmatrix} 3 &0&1 \\
0 &1&0 \\
1 &0&3
\end{pmatrix}$$
Condition of biorthogonality for example for$f^{(2)}_1$ is written as
$$  
           \left\{  
           \begin{array}{rcl}  
           (f^{(2)}_1,g^{(2)}_1) & = & 1  \\
 (f^{(2)}_1,g^{(2)}_2)      & = & 0\\
 (f^{(2)}_1,g^{(2)}_3)  & = &   0\\
           \end{array}  
\right.
  $$  
that is
$$  
           \left\{  
           \begin{array}{rcl}  
          3a_1+c_1 & = & \frac {4}{\pi}  \\
 b_1     & = & 0\\
 a_1+3c_1 & = &   0\\
           \end{array}  
\right.
  $$  
The solution is $f^{(2)}_1(x,y)=\frac{1}{2\pi }(3x^2-y^2)$. Similarly, the other elements of the system are 
$$f^{(2)}_2(x,y)=\frac {4}{\pi }xy, \;\;\; f^{(2)}_3(x,y)=\frac{1}{2\pi }(-x^2+3y^2)$$
\\
\par
For $n=3$ we have 
$$g^{(3)}_1(x,y)=x^3, \;\;g^{(3)}_2(x,y)=x^2y, \;\;g^{(3)}_3(x,y)=xy^2, \;\;g^{(3)}_4(x,y)=y^3$$
Gram matrix in the basis of $g^{(3)}_s(x,y)$ is
$$\Gamma = \frac{\pi}{8}\begin{pmatrix} 5 &0&1&0 \\
0 &1&0&1 \\
1 &0&1&0 \\
0&1&0&5\\
\end{pmatrix}$$
By similar calculations we obtain

$$f^{(3)}_1(x,y)=\frac{2}{\pi}(x^3-xy^2)$$

$$f^{(3)}_2(x,y)=\frac{2}{\pi}(5x^2y-y^3)$$

$$f^{(3)}_3(x,y)=\frac{2}{\pi}(5xy^2-x^3)$$

$$f^{(3)}_4(x,y)=\frac{2}{\pi}(y^3-x^2y)$$

\end{document}